\documentclass[12pt]{article}

\usepackage[T2A]{fontenc}
\usepackage[cp1251]{inputenc}
\usepackage[english,russian]{babel}
\usepackage[tbtags]{amsmath}
\usepackage{amsfonts,amssymb}

\usepackage{mathrsfs}
\usepackage[paper]{mz2ewin}

\numberwithin{equation}{section}

\theoremstyle{plain}

\theoremstyle{definition}

\newcommand{\const}{\mathrm{const}}

\begin{document}

\normalfont\color{black}

\title{Nonlinear Maxwell equations}
\author{Sergey Y. Kotkovskiy}
\email{s\_kotkovsky@mail.ru, s\_kotkovsky@yahoo.com}
\date{February 29, 2024}

\udk{537.8}

\markboth{S.\,Y.~Kotkovskiy}{Nonlinear Maxwell equations}

\maketitle

\begin{fulltext}

\begin{abstract}
\begin{footnotesize}
Based on the analysis of biquaternion quadratic forms of field, it is shown that Maxwell equations arise as a consequence of the principle of conservation of the energy-momentum flow of field in space-time. It turns out that this principle presupposes the existence
more general nonlinear field equations. Classical linear Maxwell equations are embedded in a special way into new nonlinear equations
and are their special case.
It is shown that in a number of important cases nonlinear equations, in contrast to linear ones,
allow solutions that have a swirling energy flow.
Solutions of the equations we obtained make it possible to give wave description of charged particles, common for quantum mechanics, within the framework of nonlinear classical electrodynamics. Special attention in the work is paid to the problem of dividing the field into "own"\  field of a charged particle
and a field "external" \  to it.
From the nonlinear field equations follow both the classical Maxwell equations themselves and the equations of charges moving under
the Lorentz force.
In this way, the problem of finding nonlinear field equations that include interaction is solved.
In our approach, the particle charge is electromagnetic (complex-valued), passing through periodically changing linear combinations of electric and magnetic charges - from purely electric to purely magnetic.
In real processes, it is not the particle charge itself that plays a role, but its phase relationship with other charges and fields.
\end{footnotesize}
\end{abstract}

\begin{flushleft}
\begin{scriptsize}
\end{scriptsize}
\end{flushleft}

\newpage

\vspace{3mm}
Contents
\vspace{3mm}

\textbf{1. Introduction.}

\textbf{2. Biquaternionic Lorentz transformations.}

\textbf{3. Nonlinear equations of free field.}

\textbf{4. Energy representation.}

\textbf{5. Derivation of nonlinear equations from the energy-momentum tensor.}

\textbf{6. Structure of energy flow.}

\textbf{7. Regular currents.}

\textbf{8. Classical Maxwell equations as a linear limit.}

\textbf{9. Associated fields.}

\textbf{10. Nonlinear plane wave.}

\textbf{11. Twisting energy flow.}

\textbf{12. Central-symmetric field. Electromagnetic charge.}

\textbf{13. Equations with one singularity.}

\textbf{14. Coherent fields.}

\textbf{Conclusions.}

\textbf{Appendix 1. Algebra of biquaternions.}

\textbf{Appendix 2. Ostrogradsky-Gauss biquaternionic formulas.}

%%%%%%%%%%%% 1. Introduction
\section{Introduction}\label{Intro}
Maxwell equations rightfully occupy one of the central places in classical and modern physics. However, until now they have remained
in the status of a postulate generalizing experimental data on the phenomena of electromagnetism.
No derivation of these equations from more fundamental principles was given.
The goal facing us here is the axiomatic derivation of the electromagnetic field equations
based on the basic principle of conservation of energy-momentum field flow in space and time.

Classical, and after it quantum electrodynamics, consider elementary charges (electrons and positrons) and the electromagnetic field (hereinafter \textit{field}) as two interacting but separate entities. The study of the behavior of interacting charged particles and fields is conducted by
joint consideration of Maxwell equations for the field generated by charges and the Lagrangian
interaction of charged particles and field. At the same time, many authors have long proposed to consider charges as singularities
fields rather than as separate objects (e.g. \cite{Lanczos}).
We also base our approach on the assumption that charged particles represent field singularities,
Moreover, the initial requirement is that only through these singularities can the field lose or gain energy-momentum.

To describe the field and its Lorentz structure, we use the mathematical apparatus of biquaternions.
Biquaternions describe Lorentz transformations in a simple and convenient form of bilinear operators \cite{Silberstein}. The electromagnetic field is represented as a three-dimensional complex-valued vector which is sometimes called Riemann-Silberstein vector.

%Imaeda
The greatest success, in our opinion, in applying biquaternion analysis to the description of the electromagnetic field was achieved by K. Imaeda, who gave a new formulation of electrodynamics based on the biquaternion analyticity introduced by him \cite{Imaeda}. The conceptual step taken by this researcher was
to affirm the unity of the electromagnetic field and Minkowski space-time. Maxwell equations are represented
not as an independent postulate, but are a consequence of the Lorentz structure of field. In his work, Imaeda examines the functional of the quadratic form formed by the biquaternion field function and the associated functional derivative.
The conditions for the regularity of the field function give simultaneously both Maxwell equations and the equation for the Lorentz force acting on a charge in this field. In the present study, we follow Imaeda's method of studying hypersurface integrals of the quadratic form of the field,
but at the same time we are based on other initial principles and arrive at intrinsically new results.

%%%Nonlinear ED
The absence of interaction of charges in the composition of Maxwell equations themselves, their not closeness, is one of the main challenges for the search for nonlinear equations of the electromagnetic field \cite{Chern},\cite{Delf}. Indeed, the linearity of the equations
assumes the independence of each of its solutions, since the sum of the solutions automatically becomes a new solution.
Each of the solutions does not affect the other; in other words, linear field equations do not describe interaction of fields.
Usually the interaction is introduced separately and comes up as Coulomb interaction or Lorentz force.
To solve this dilemma, A.A. Chernitsky proposed to use the nonlinear Born-Infeld model \cite{Chern}.
In this direction, he obtained a number of interesting solutions, such as field solitons.
We consider the Born-Infeld equations as the first attempt at a nonlinear generalization of Maxwell equations.

In our approach, we follow the same ideology of a unified nonlinear field, in which charged particles are special singular configurations of this field, but we do not resort to a model, but derive equations from universal principles. We start from the fundamental physical principle
of conservation of energy-momentum, according to which as much of it flows into a given 4-volume, so much should flow out [6].
We are looking for a description of the movement of this "river" \  of energy-momentum in Lorentz-covariant field structures in their biquaternion representation.
Obtaining nonlinear equations on this way happened to be a pleasant surprise.
At the same time, the beauty of Maxwell classical linear equations is not lost - they turn out to be
embedded in the new nonlinear equations.

Section \ref{Base} briefly introduces the reader to the mathematical apparatus and the basics of the approach.
Sections \ref{Free}-\ref{nelin_volna} are dedicated to
derivation and analysis of nonlinear free field equations.
In section \ref{tensor} we present the derivation of these equations in the more conventional
tensor form.
In sections \ref{centr}-13
particle nonlinear fields and their interactions are studied.

%%%%%%%%%%%% Key points
\section{Biquaternionic Lorentz transformations}\label{Base}
Space-time is represented by a real biquaternion space\footnote{We use a system of units in which the speed of light is equal to one: $c=1$.} $Z=(t,\bold r)$. Basic operations with biquaternions are given in Appendix 1 (also see \cite{Berezin}).
%Lorentz transformation
General Lorentz transformations of space-time, including boosts and rotations, have the form of a biquaternion product \cite{Silberstein}:
\begin{equation}
\label{PL_koord}
Z' = U^\ast Z U,
\end{equation}
%%%
where $U$ is some unitary biquaternion: $|U|^2=1$. $U$ is represented as:
\begin{equation}
U=e^{\boldsymbol \theta} = \ch{\theta} + \frac{\boldsymbol \theta}{\theta} \sh{\theta} ,
\end{equation}
where $\boldsymbol \theta$ is an arbitrary three-dimensional complex vector of nonzero value $\theta$ ($\boldsymbol \theta^2 \ne 0$).
Real $\boldsymbol \theta$ define special Lorentz transformations (boosts), while imaginary $\boldsymbol \theta$ define spatial rotations.

In particular, boost at a speed of $\bold V$, i.e. the transition to a reference frame moving at a speed of $(-\bold V)$ has the form of a biquaternion:
\begin{equation}
\label{bust}
U = ( \sqrt{\frac{\gamma+1}{2}}, \frac{\bold V}{V} \sqrt{\frac{\gamma-1}{2}} ),
\end{equation}
where the standard denotation is $\gamma = \frac{1}{\sqrt{1-V^2}}$. The $\boldsymbol \theta$ parameter is related to the boost speed by the relation: $\bold V = \th{2 \boldsymbol \theta} = \frac{\boldsymbol \theta}{\theta} \th 2\theta$.

Let's consider the motion of a particle of some mass.
Its energy-momentum $P=(\epsilon,\bold p)$ is transformed similarly to $Z$, i.e. is a \textit{4-vector}:
\begin{equation}
\label{PL_imp}
P' = U^\ast P U
\end{equation}

Over $Z$ you can introduce a complex vector field that describes the electromagnetic field (Riemann-Silberstein vector) \cite{Bial} (p. 8-9):
\begin{equation}
\bold F=\bold F(Z) = \bold E+ i \bold H,
\end{equation}
whose components $\bold E$ and $\bold H$ are the electric and magnetic field strengths, respectively.\footnote{Here and below, any function of the variable $Z$ includes a dependence on both $Z$ and $\overline Z $, in other words means a function of the variables $t$ and $\bold r$.}
Lorentz transformations of the field $\bold F$ have the form:
\begin{equation}
\label{PL_pole}
\bold F' = \overline U \bold F U,
\end{equation}
where $\overline U$ denotes the biquaternion conjugate to the biquaternion $U$. For $U = (s,\bold u)$: $\overline U = (s, - \bold u)$.

The above statements about the Lorentz structure of field and particles,
together with the central principle - the law of conservation of energy-momentum in space-time, as well as the requirement that energy be quadratic over the field,
constitute the axiomatics of our approach.

Let us briefly describe our method.
First, based on the complex vector field $\bold F(Z)$, we find the form of a real biquaternion quadratic form,
which behaves under Lorentz transformations like energy-momentum.
Such a structure is defined at each point in space-time and is expressed as the differential of some biquaternion function $\Sigma(Z)$, which has the meaning of the energy-momentum flow of the field. In the tensor representation of the function $\Sigma(Z)$ it corresponds the field energy-momentum tensor.

Then, we formulate the principle of conservation of $\Sigma(Z)$ in space-time, which is expressed in the equality to zero of the integral
from $d\Sigma$ along any closed hypersurface of four-dimensional pseudo-Euclidean space. Using the biquaternion version of the Ostrogradsky-Gauss formula, the last integral
is replaced with the integral over the space-time region surrounded by a given hypersurface. As a result, we obtain differential field equations, which represent a nonlinear generalization of Maxwell equations.

%%%%%%%%%%%% Free field equations

\section{Nonlinear equations of free field}\label{Free}
\textit{Singularities} of the field are its parts which turn to infinity inside a finite space-time volume.
In this section we will derive the equations of \textit {free field} (or fields in "emptiness"), i.e. a field that does not have singularities in a given region of space-time.
Let us clarify that our definition of a free field is not the same as the definition of a free field in classical linear electrodynamics,
where a free field means a field that is not associated with any source.

The energy-momentum of a free field must be completely conserved in space-\linebreak-time, which means it should not transform into other types of energy, as is the case in the presence of singularities that play the role of channels for the exchange of energy between different types of motion.

Let us consider some space-time hypersurface (three-dimensional manifold) $S_3$.
For any element of this hypersurface one can define a biquaternion $dZ$
orthogonal to it, the magnitude of which is equal to the 3-volume of the hypersurface element\footnote{External differential forms of quaternions are studied in detail in the work [14]. A generalization to the case of real biquaternions is given in the work \cite{Imaeda}.}.

For the field $\bold F$ on an arbitrary section of the hypersurface, we can define a real-valued quadratic form that behaves under Lorentz transformations similar to the energy-momentum of a particle (\ref{PL_imp})\footnote{The factor $\frac{1}{2}$ is introduced here to be consistent with the generally accepted definition of the Umov-Poynting vector (see below (\ref{Umov2})).}:
\begin{equation}
\label{sigma}
d \Sigma = \frac{1}{2} \bold F^{\ast} dZ \bold F
\end{equation}
Indeed, in accordance with (\ref{PL_koord}), $dZ$ is transformed as
\begin{equation}
\label{sigma2}
dZ' = U^\ast dZ U
\end{equation}
and taking into account (\ref{PL_pole}) the value $d \Sigma$ is transformed as
\begin{equation}
\label{PL_sigma}
d \Sigma' = \frac{1}{2} U^\ast \bold F^\ast \overline{U^\ast} U^\ast dZ U \overline U \bold F U = U ^\ast \frac {1}{2} \bold F^\ast dZ \bold F U
\end{equation}
or
\begin{equation}
\label{PL_sigma2}
d\Sigma' = U^\ast d\Sigma U,
\end{equation}
which coincides in form with the Lorentz energy-momentum transformation (\ref{PL_imp}).

You can directly verify that, apart from (\ref{sigma}), there are no other real-valued quadratic forms of the field that transform like energy-momentum\footnote{see. Appendix\ref{Appendix_SP}}. For example, a form  $d \Sigma = \frac{1}{2} \bold F \bold F^{\ast} dZ$ does not have this property.

The value $\Sigma$ has the meaning of the energy-momentum flux of the field, which includes both the spatial densities of energy and momentum of the field, and their fluxes in space. For brevity, we will also call $\Sigma$ \textit{energy flow} of the field. The formula (\ref{sigma}) connects three fundamental quantities: energy, field and space-time. According to (\ref{sigma})
energy is factorized by the field, thereby providing space-time filling. The structure of the energy flow $d \Sigma$ is considered in more detail in the Appendix \ref{Appendix_SP}.

We require the continuity (conservation) of the $\Sigma$ flow in space-time.
This will mean that the total flow $\Sigma$ through any closed space-time hypersurface $S_3$ is equal to zero:
\begin{equation}
\label{sigma_int}
\oint_{S_3} \,d \Sigma = \frac{1}{2} \oint_{S_3} \bold F^{\ast} dZ \bold F = 0
\end{equation}

According to the Ostrogradsky-Gauss formula, under the condition of continuous coordinate differentiability of the components of the function $\bold F$, the integral on the left side of the equation (\ref{sigma_int}) is represented as an integral over the four-dimensional space-time volume $V_4$, surrounded by the hypersurface $ S_3$ (see Appendix \ref{Appendix_OG}):
\begin{equation}
\label{sigma_int2}
  \oint_{S_3} \bold F^{\ast} dZ \bold F = \int_{V_4} ( \bold F^{\ast} D \bold F) dV_4
\end{equation}
where $(\bold F^{\ast} D \bold F)$ denotes the quadratic form:
\begin{equation}
\label{kv_forma}
( \bold F^{\ast} D \bold F) = 2 Re\{ \bold F^{\ast} ( D \bold F) \} = \bold F^{\ast} ( D \bold F) + ( \bold F^{\ast} D ) \bold F,
\end{equation}
a $D = (\partial_t, \nabla )$ is a biquaternion gradient, the action of which on some biquaternion $(s,\bold u)$ on the left and right is expressed according to the rules of biquaternion product (\ref{refText3}) as:

\begin{equation}
\label{grad1}
   D (s,\bold u) = (\partial_t, \nabla ) (s,\bold u) =
( \partial_t s + \nabla \cdot \bold u, \ \partial_t \bold u + \nabla s + i \nabla \times \bold u )
\end{equation}
\begin{equation}
\label{grad2}
  (s,\bold u) D = (s,\bold u) (\partial_t, \nabla ) =
( \partial_t s + \nabla \cdot \bold u, \ \partial_t \bold u + \nabla s - i \nabla \times \bold u )
\end{equation}
respectively.
The equation (\ref{sigma_int}) will be written as:
\begin{equation}
\label{sigma_int3}
  \int_{V_4} ( \bold F^{\ast} D \bold F ) dV_4 = 0
\end{equation}
Since this condition must be satisfied for any closed 4-volume $V_4$, then at any point in space-time we have:
\begin{equation}
\label{um1}
( \bold F^{\ast} D \bold F ) = 0
\end{equation}
or
\begin{equation}
\label{um2}
Re\{ \bold F^{\ast} ( D \bold F) \} = 0
\end{equation}
We can write the last equation in the form:
\begin{equation}
\label{RUM10}
\begin{cases}
  Re\{ (\partial_t \bold F) \cdot \bold F^{\ast} - i [\nabla \bold F] \cdot \bold F^{\ast} \} \ = \ 0 \\
Re\{ ( \nabla \bold F ) \bold F^{\ast} + i (\partial_t \bold F) \times \bold F^{\ast} + [\nabla \bold F] \times \bold F ^{\ast} \} \ = \ 0 \\
\end{cases}
\end{equation}
or in more detail:
\begin{equation}
\label{RUM}
\begin{cases}
\bold E \cdot ({\partial}_t \bold E - \nabla \times \bold H) + \bold H \cdot ({\partial}_t \bold H + \nabla \times \bold E) = 0 \ \
\bold E (\nabla \cdot \bold E) + \bold H (\nabla \cdot \bold H) = \bold E \times ({\partial}_t \bold H + \nabla \times \bold E) - \bold H \times ({\partial}_t \bold E - \nabla \times \bold H)
\end{cases}
\end{equation}
The equations (\ref{RUM}) are \textit{Maxwell nonlinear equations for a free field}.
By virtue of their construction, these equations are Lorentz-covariant: they are preserved when moving to any other reference system.
Each of the equations (or systems of equations) (\ref{um1}),(\ref{um2}),(\ref{RUM10}) is equivalent to the equations (\ref{RUM}), thereby being alternative formulations of Maxwell nonlinear equations.

%%%%%%%%%%%% Energy representation
\section{Energy representation}\label{EnergPredst}

The equations (\ref{RUM}) can also be represented as:
\begin{equation}
\label{energ}
\begin{cases}
\frac{1}{2} \partial_t (E^2 + H^2) + \nabla \cdot (\bold E \times \bold H) = 0\\
\partial_t (\bold E \times \bold H) = ( \bold E \cdot \nabla) \bold E + ( \bold H \cdot \nabla) \bold H
\end{cases}
\end{equation}
The last equations involve
spatial energy density of the field $W$ and its flux $\bold S$ (momentum density, or Umov-Poynting vector), together forming a biquaternion of energy-momentum density:
\begin{equation}
P = (W,\bold S) = \frac{1}{2} \bold F \bold F^{\ast}
\end{equation}
\begin{equation}
\label{Umov}
\begin{cases}
W = \frac{1}{2}\bold F \cdot \bold F^{\ast} = \frac{1}{2} (E^2 + H^2) \\
\bold S = \frac{i}{2} \bold F \times \bold F^{\ast} = \bold E \times \bold H
\end{cases}
\end{equation}
Using $W,\bold S$ (\ref{energ}) are represented in \textit{energy form}:
\begin{equation}
\label{Umov2}
\begin{cases}
  \partial_t W + \nabla \cdot \bold S = 0\\
  \partial_t \bold S + \nabla W = \bold Q,
\end{cases}
\end{equation}
where indicated:
\begin{equation}
\bold Q =\big ( (\nabla \cdot \bold E) + \bold E \cdot \nabla\big) \bold E +
\big( (\nabla \cdot \bold H) + \bold H \cdot \nabla\big) \bold H
\end{equation}

The first of the equations (\ref{Umov2}) is the equation of conservation (continuity) of the field energy density, and the second is the equation of its momentum density.
As shown in the next section, these equations are equivalent to the energy-momentum continuity equations following from the classical field energy-momentum tensor.

Note that for ordinary linear Maxwell equations there is no energy representation similar to (\ref{Umov2}).
The presence of the latter is, therefore, an exclusive property of nonlinear equationsfield definition.

%%%%%%%%%% Energy-momentum tensor %%%%%%%%%%%
\section{Derivation of nonlinear equations from the energy-momentum tensor}\label{tensor}
According to Sommerfeld, the field energy-momentum tensor has a more direct relationship to physical reality than the field quantities themselves, and the equations associated with this tensor are more fundamental in nature than Maxwell equations \cite{Strazh} (p. 64).

The field energy-momentum tensor $T^{ik}$ has five independent components, while the field itself has six, which indirectly indicates
about the presence of one more degree of freedom. This allows you to enter the so-called \textit{dual rotations} \cite{Strazh}, which come down to the multiplication of all electromagnetic fields by some constant phase factor, the same everywhere and for all fields.
Dual rotations lead to the appearance in the theory of magnetic charges along with existing electric ones. As it turns out, such a theory describes the same observed interaction effects of
fields and charged particles, as in the theory, which operates only with electric charge.

In section 14 we will find out that the dual rotation does not have to be global - constant throughout space-time and for all particles.
In the case of two particles we study below, in order for the theory to correspond to the existing observed effects, only a certain coordination of the dual rotations of the fields of these particles is sufficient.
Using dual rotations, both the well-known global ones and the local ones introduced in this paper below,
we will see that the classical field energy-momentum tensor allows a wider class of interacting fields than is generally accepted.

We will show now the equivalence of the biquaternion energy-momentum flow $\Sigma$,
introduced by us in the section \ref{Free},
to the classical energy-momentum tensor $T^{ik}$.
To do this, we present another derivation of the nonlinear field equations (\ref{RUM}) - from the classical energy-momentum tensor. Unlike the main part of the article, in this section we follow the tensor representation of the field and terminology \cite{LL}\footnote{Our units of measurement of field quantities differ from those adopted in \cite{LL} by a factor of $\frac{1}{4 \pi }$, as well as in our system of units
speed of light $c=1$.}.

The field energy-momentum tensor has the form \cite{LL}(32.15):

\begin{equation}
\label{tensor1}
T^{ik} =
\begin{bmatrix}
     W & S_x & S_y & S_y \\
    S_x & -\sigma_{xx} & -\sigma_{xy} & -\sigma_{xy} \\
    S_y & -\sigma_{yx} & -\sigma_{yy} & -\sigma_{yz} \\
    S_z & -\sigma_{zx} & -\sigma_{zy} & -\sigma_{zz} \\
\end{bmatrix}
\end{equation}
where $W$ and $\bold S$ are the energy and momentum densities of the field, defined above in (\ref{Umov}),
and $\sigma_{\alpha \beta}$ is the Maxwell stress tensor \cite{LL}(33.3):
\begin{equation}
\label{tensor2}
\sigma_{\alpha \beta} = E_\alpha E_\beta + H_\alpha H_\beta - \frac{1}{2} \delta_{\alpha \beta} (E^2 + H^2)
\end{equation}
In the equations (\ref{tensor1}),(\ref{tensor2}), the Latin indices are spatiotemporal and range from 0 to 3, while the Greek indices are purely spatial and range from 1 to 3.

The energy-momentum continuity equations are obtained by applying the divergence operation to the tensor $T^{ik}$ \cite{LL}(32.4):
\begin{equation}
\label{tesor0}
\frac{\partial T^{k}_{i}}{\partial x^k} = 0
\end{equation}
Dividing this equation into spatial and temporal components instead, we obtain  two other equations (\cite{LL} (32.12)):
\begin{equation}
\label{tensor_nepr1}
\frac{\partial T^{00}}{\partial t} \ + \ \frac{\partial T^{0\alpha}}{\partial x^\alpha} = 0
\end{equation}
\begin{equation}
\label{tensor_nepr2}
\frac{\partial T^{\alpha0}}{\partial t} \ + \ \frac{\partial T^{\alpha\beta}}{\partial x^\beta} = 0
\end{equation}

According to the values of the components of the tensor $T^{ik}$ in (\ref{tensor1}), the equation (\ref{tensor_nepr1}) takes the form:
\begin{equation}
  \partial_t W + \nabla \cdot \bold S = 0
\end{equation}
which coincides with the first of our equations (\ref{Umov2}).

Now we can move on to the second of the continuity equations - the equation (\ref{tensor_nepr2}).
Take for example this equation for the $x$ coordinate:
\begin{equation}
\label{partp}
  \frac{\partial S_x}{\partial t} - \nabla \cdot \boldsymbol \chi = 0,
\end{equation}
where $\boldsymbol \chi$ is the following vector:
\begin{equation}
\boldsymbol \chi =
\begin{pmatrix} \sigma_{xx} \\ \sigma_{xy} \\ \sigma_{xz} \end{pmatrix} =
  \begin{pmatrix} \frac{1}{2}(E^2_x - E^2_y -E^2_z + H^2_x - H^2_y -H^2_z) \\ E_x E_y + H_x H_y \\ E_x E_z + H_x H_z\end{pmatrix}
\end{equation}
Further,
\begin{equation}
\label{partp2}
\begin{split}
  \nabla \cdot \boldsymbol \chi =
  E_x (\partial_x E_x + \partial_y E_y + \partial_z E_z ) + E_y( \partial_y E_x - \partial_x E_y) + E_z( \partial_z E_x - \partial_x E_z) + \\
  +
H_x (\partial_x H_x + \partial_y H_y + \partial_z H_z ) + H_y( \partial_y H_x - \partial_x H_y) + H_z( \partial_z H_x - \partial_x H_z)
\end{split}
\end{equation}
The equation (\ref{partp}) becomes:
\begin{equation}
\label{sx}
\begin{split}
  \frac{\partial S_x}{\partial t} =
  E_x (\partial_x E_x + \partial_y E_y + \partial_z E_z ) + E_y( \partial_y E_x - \partial_x E_y) + E_z( \partial_z E_x - \partial_x E_z) + \\
  +
H_x (\partial_x H_x + \partial_y H_y + \partial_z H_z ) + H_y( \partial_y H_x - \partial_x H_y) + H_z( \partial_z H_x - \partial_x H_z)
\end{split}
\end{equation}
Let's make sure that this equation coincides with the second of our equations (\ref{Umov2}) for the $x$-component.
Indeed:
\begin{equation}
(\nabla W)_x = E_x \partial_x E_x + E_y \partial_x E_y + E_z \partial_x E_z +
E_x \partial_x E_x + E_y \partial_x E_y + E_z \partial_x E_z
\end{equation}
\begin{equation}
\begin{split}
Q_x = ( \partial_x E_x + \partial_y E_y +\partial_z E_z + E_x \partial_x + E_y \partial_y + E_z \partial_z ) E_x + \\
+( \partial_x E_x + \partial_y E_y +\partial_z E_z + E_x \partial_x + E_y \partial_y + E_z \partial_z ) E_x
\end{split}
\end{equation}
which directly implies the coincidence of the second of the equations (\ref{Umov2}) and the equation (\ref{sx}) for the $x$-components.
It is obvious that these equations coincide for both $y$ and $z$ components, and, therefore, coincide in general.
We have verified that the second of the tensor continuity equations (\ref{tensor_nepr2}) is equivalent to the second of the equations (\ref{Umov2}):
\begin{equation}
  \partial_t \bold S + \nabla W = \bold Q
\end{equation}

So, we have shown that the energy equations (\ref{Umov2}) are equivalent to the energy-momentum continuity equations (\ref{tensor_nepr1}), (\ref{tensor_nepr2}) associated with the classical energy-momentum tensor. But this means the equivalence of Maxwell nonlinear equations (\ref{RUM}) to the tensor energy-momentum continuity equation (\ref{tesor0}), which is what we wanted to prove.

%%%%%%%%%%%% thread structure
\section{Structure of energy flow. Opposite direction in time of particle and field energy flows.}\label{Appendix_SP}

The structure of the field energy-momentum flow is studied in detail, based on differential forms, in the book \cite{Wiler} (p. 174-181).
Here we will give a simple diagram that gives a basic understanding of this structure.
The energy flow $d \Sigma$ passing through the oriented three-dimensional space-time hypersurface $d S_3$ is given by the formula (\ref{sigma}):
\begin{equation}
\label{sigma_2}
d \Sigma = \frac{1}{2} \bold F^{\ast} dZ \bold F
\end{equation}
The "platform" \ $d S_3$ is expressed by the biquaternion $dZ$.
By definition, the scalar part $d \Sigma $ is equal to the energy flow, and the vector part is equal to the momentum flow. Figure \ref{Ris_potoki} conventionally depicts two special cases of flow $d \Sigma$, which will be discussed below. The abscissa is space (schematically three spatial dimensions are combined into one), and the ordinate is time.

In the first case, the area $dZ_V$ is a certain three-dimensional volume "oriented" \ forward in time, and $d \Sigma_T$ is a timelike flow.
\begin{equation}
\label{str_T}
d \Sigma_T = \frac{1}{2} \bold F^{\ast} dZ_V \bold F = (W dt, -\bold S dt)
\end{equation}
The hyperplatform $d Z_T$, expressed in terms of the directional time element $dt$, is orthogonal to all spatial dimensions.
Continuity of energy flow in this case means conservation of energy and momentum over time.
The "area" \ of the hyperplatform $d Z_T$ (denoted by the value $dt$!) is equal to the three-dimensional volume $dV$ through which the flow passes.
This confirms that the quantities $W$ and $-\bold S$ express the spatial densities of energy and momentum of the field.

$d \Sigma_T$ is a "vertical" \ energy-momentum flow reversed in time, which is expressed by the minus sign in front of the Umov-Poynting vector $\bold S$.
The time-reverse nature of the field pulse-energy is due to the type of energy flow differential (\ref{sigma}). As the energy flow differential, one could choose the conjugate value (\ref{sigma}): $d \Sigma_S = \overline{ d \Sigma} = \frac{1}{2} \bold F \overline{ dZ} \bold F^{\ast}$. Such a flow of field energy would be directed along time, and not back to it. However, $d \Sigma_S$ is transformed not as a 4-momentum energy-momentum, but as its conjugate biquaternion\footnote{Biquaternion conjugate corresponds to the transition from covariant to contravariant 4-vectors in the tensor representation.}. This means that such a choice is invalid if we want the total energy-momentum value of the field and particle to retain the properties of a Lorentz 4-vector. An important conclusion follows from this: the energy of the particleв's field flows in the direction of time opposite to that in which the energy of the particle itself flows. Otherwise it can be said that time "flows" in different directions in kinematic and field \textit{planes}. This points to  existence of different than usually accepted physiclal type of causality - happening events may depend not only on past configuration but also on future one.

In the second case, the hyperplatform $d Z_S$ has as "edges" \ a time interval $dt$ and a section of an oriented two-dimensional spatial surface (vector $d \bold r$). The energy-momentum flow $d \Sigma_S$ is timelike.
\begin{equation}
\label{str_S}
d \Sigma_S = (W dt + (\bold S d \bold r), -\bold S dt+ \bold E(\bold E d \bold r) + \bold H (\bold H d \bold r) - W d\bold r)
\end{equation}%%
  The terms $W dt$ and $-\bold S dt$ have the same meaning of flows in time as in the first case.
The quantity $(\bold S d \bold r)$ determines the change in energy due to its transfer in the spatial dimension.
This corresponds to the well-known fact that the Umov-Poynting vector is the energy flux density in space.
The quantity $ \bold E(\bold E d \bold r) + \bold H (\bold H d \bold r) - W d \bold r$ gives the transfer of momentum in space. It
corresponds to the Maxwell stress tensor.

  \begin{figure}
\label{Ris_potoki}
\centering
\includegraphics[width=2.755in,height=2.12in]{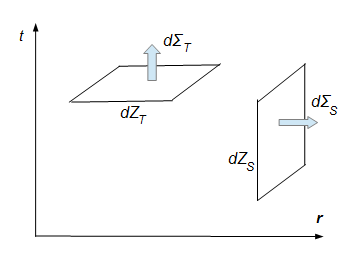}\caption{Structure of the field energy flow.}
\end{figure}
%%

%%%%%% Regular currents
\section{Regular currents}
In the nonlinear field equations we derived, a special role is played by the biquaternion value of the \textit{4-current} of a given field:
\begin{equation}
\label{bikv_tok}
\mathscr{J} \equiv D \bold F = ( \nabla \cdot \bold F, \partial_t {\bold F} + i \nabla \times \bold F)
\end{equation}
For a free field, the 4-current, like the field itself, is regular (not singular)
function of the space-time coordinate $Z$.
We will call such a non-singular current \textit{regular current}.
$\mathscr{J}$ is some regular function of $Z$ defined by the field $\bold F$ itself.

The 4-current is decomposed as the sum of complex charge and current:
\begin{equation}
\label{bikv_tok2}
\mathscr{J} = ( e_0 + i g_0, \bold J_0 + i \bold I_0 )
\end{equation}
$e_0$ and $\bold J_0$ are regular electric charge and current, $g_0$ and $\bold I_0$ are regular magnetic charge and current, defined as:
\begin{equation}
\label{kvazi_el_tok}
\begin{cases}
e_0 = \nabla \cdot \bold E\\
g_0 = \nabla \cdot \bold H \\
\bold J_0 = {\partial}_t \bold E - \nabla \times \bold H \\
\bold I_0 = {\partial}_t \bold H + \nabla \times \bold E
\end{cases}
\end{equation}

In terms of the regular 4-current, the nonlinear equations (\ref{um1}) will be written as:
\begin{equation}
  Re\{ \bold F^{\ast} \mathscr{J} \} = 0,
\end{equation}
or in expanded form:
\begin{equation}
\label{RUM2}
\begin{cases}
\bold J_0 \cdot \bold E + \bold I_0 \cdot \bold H = 0\\
e_0 \bold E + g_0 \bold H - \bold J_0 \times \bold H + \bold I_0 \times \bold E = 0
\end{cases}
\end{equation}

Note that our definitions (\ref{kvazi_el_tok}) coincide in form with the classical Maxwell equations with the right side in symmetric electrodynamics,
allowing charges and currents of both electric and magnetic types \cite{Strazh}.
In addition, it is easy to verify that the continuity condition is satisfied for regular charges and currents:
\begin{equation}
\label{nepr1}
\begin{cases}
\frac{\partial e_0}{\partial t} + \nabla \cdot \bold J_0 = 0\\
\frac{\partial g_0}{\partial t} - \nabla \cdot \bold I_0 = 0
\end{cases}
\end{equation}

Ordinary (point) charges and currents have singular character and therefore are fundamentally different from regular charges and currents.
However, using the similarity of regular and ordinary charges and currents, one can see in the equations (\ref{RUM2}) the fact that the free field does not apply work to its regular currents (the first equation), and that the total force applied from the field to its regular charges and currents is equal to zero (second equation). This is a natural consequence of the original principle of conservation of field energy. Nonlinear field equations (\ref{RUM}) are \textit{recursive} in the sense that the sources of a free field can be its own nonsingular structures - regular currents with which this field interacts without doing work on them.

It is important not to confuse the regular charges and currents we introduced with the spatial density of charges and currents of macroscopic electrodynamics\cite{LL8},
despite their apparent closeness.
In the electrodynamics of continuous media we deal with distributed point charges and currents averaged over a spatial region,
while regular charges and currents are not the result of averaging.
Further (in the Conclusions section) it is indicated that regular currents can serve as a classical analogue of vacuum currents of quantum field theory.

%%%%%%%%%% Classical Maxwell equations
\section{Classical Maxwell equations as a linear limit}\label{OUM}
Based on general considerations, if there are some nonlinear equations of the electromagnetic field, then the classical Maxwell equations should be their linear approximation \cite{Delf} (p.6). Also linear equations stop working
at sufficiently high field strengths.
In our case, this is exactly what happens within our aproach: the classical Maxwell equations in emptiness are a special case of nonlinear ones (\ref{um2}) with zero regular 4-currents:
\begin{equation}
\label{oum1}
  D\bold F = 0
\end{equation}
Indeed, using the definition (\ref{grad2}), it is easy to verify that the expanded equation (\ref{oum1}) looks like:
\begin{equation}
\label{Maxwell}
\begin{cases}
{\partial}_t \bold E - \nabla \times \bold H = 0 \\
{\partial}_t \bold H + \nabla \times \bold E = 0 \\
\nabla\cdot \bold E = 0 \\
\nabla \cdot \bold H = 0
\end{cases}
\end{equation}

Nonlinear equations (\ref{RUM}) always hold when classical Maxwell equations (\ref{Maxwell}) are satisfied, but the reverse is not true. Examples of nonlinear fields,
for which nonlinear equations are satisfied, but the usual Maxwell equations are not satisfied, are given below in section \ref{vert}.

Now we will go along how the linear approximation of our equations  stops working at high field strengths.
Nonlinear equations (\ref{um2}) are linearized at sufficiently small values of the total current:
\begin{equation}
| D\bold F| < \delta
\end{equation}
where $\delta$ is some small value determined by the accuracy of a particular observation; module assessment
refers to both the scalar and vector components of the current.
The equations (\ref{RUM}) are approximately satisfied at such field values at which:
\begin{equation}
|\bold F^{\ast} ( D \bold F ) | \ll 1 \ \Leftrightarrow |\bold F| \ll \frac{1}{\delta}
\end{equation}
from which it follows that for field strength values $|\bold F| >\frac{1}{\delta}$
linear approximation ceases to be valid.

%%%%%%%%%% Related fields
\section{Associated fields}\label{vert}
We will call the solutions to nonlinear equations (\ref{RUM}), which in the general case are not solutions to linear equations (\ref{Maxwell}), \textit{nonlinear fields}. In most cases, the sum of two or more nonlinear fields is no longer
solving field equations (\ref{RUM}).

Let $\bold F_0$ be some solution of linear equations (\ref{Maxwell}), and $(\omega_1, \bold k_1)$ some 4-vector. By using
the latter, you can construct a scalar function:
\begin{equation}
\label{scal_func}
f(t, \bold r) = e^{i (\omega_1 t - \bold k_1 \cdot \bold r)}
\end{equation}
Then
\begin{equation}
\label{nelin_resh}
\bold F= \bold F_0 f(t,\bold r) = \bold F_0 e^{i (\omega_1 t - \bold k_1 \cdot \bold r)}
\end{equation}
there is a solution to nonlinear equations (\ref{RUM}). Thus, nonlinear fields can be obtained from linear ones by multiplying the latter
to the Lorentz-invariant phase wave factor $f(t, \bold r)$.
We will call the nonlinear field $\bold F$ constructed in this way
\textit{associated} the linear field $\bold F_0$.
Inside the wave phase in (\ref{nelin_resh}), you can also take the "plus" \ sign, but the effects associated with this will be important only later - when considering
fields of singularities, so here for simplicity we limit ourselves to the "minus" \ sign.
Note that there is a more general form of solutions of the associated type than (\ref{nelin_resh}):
$\bold F= \bold F_0 e^{i \Phi(t,\bold r)}$, where $\Phi(t,\bold r)$ is some Lorentz invariant.

The field $\bold F_0$ satisfies the linear Maxwell equations (\ref{oum1}): $D \bold F_0 = 0$.
After simple calculations we get:
\begin{equation}
\label{soput}
D \bold F = ( \bold F_0 \cdot \nabla f, \ \bold F_0 \ \partial_t f - i \bold F_0 \times \nabla f),
\end{equation}
where we learned. that $\partial_t f = i \omega f$, $\nabla f = -if \bold k$, $f f^{\ast}=1$.
Nonlinear equations for the field $\bold F$ in the form (\ref{RUM10}) give:
\begin{equation}
\label{primer}
\begin{cases}
Re\{ i \omega_1 (\bold F_0 \bold F_0^{\ast}) + [\bold F_0 \bold F_0^{\ast}] \cdot \bold k_1 \} = 0\\
Re\{i \bold k (\bold F_0 \bold F_0^\ast) + \omega [\bold F_0 \bold F_0^\ast] -i (\bold F_0 \bold k_1) \bold F_0^{\ast} - i (\bold F_0^{\ast} \bold k_1) \bold F_0 \} = 0
\end{cases}
\end{equation}
Taking into account the real value character of the quantities $\frac{1}{2}\bold F \cdot \bold F^{\ast}$, $\frac{i}{2} \bold F \times \bold F^{\ast} $
and $(\bold F_0 \bold k_1) \bold F_0^{\ast} + (\bold F_0^{\ast} \bold k_1) \bold F_0$
both of these equations are obviously satisfied.

From (\ref{soput}) it follows $D \bold F \ne 0$ . This means that the field $\bold F$ (\ref{nelin_resh}) in general case does not satisfy the usual Maxwell equations (\ref{oum1}) and, therefore, is a nonlinear field.

Based on (\ref{soput}) the 4-current of associated field has the form:
\begin{equation}
\label{soput_tok}
\mathscr{J}= D \bold F = f \big(-i (\bold F_0 \bold k_1), i \omega_1 \bold F_0 + \bold [\bold F_0 \bold k_1] \big) = i f \bold F_0 \overline K_1
\end{equation}
%%

%%%%%%%%%% Nonlinear plane wave
\section{Nonlinear plane wave}\label{nelin_volna}
We take as the initial classical field $\bold F_0$ in (\ref{nelin_resh}) a linear plane wave
circularly polarized
\begin{equation}
\label{plosk1}
\bold F_0 = \bold A e^{i (\omega_0 t - \bold k_0 \cdot \bold r)} , \ \bold A^2 = 0,\ \bold A=const
\end{equation}
The wave 4-vector of a linear wave (\ref{plosk1}) $(\omega_0,\bold k_0)$ is isotropic: $\omega_0^2 = \bold k_0^2$, and this wave is transverse: $\bold k_0 \perp  \bold A$.
An associated solution (\ref{plosk1}) with phase factor $f = e^{i (\omega_1 t - \bold k_1 \cdot \bold r)}$ has the form:
\begin{equation}
\label{plosk_RUM}
\bold F= f \bold F_0 = \bold A e^{i (\omega t - \bold k \cdot \bold r)}
\end{equation}
Where
\begin{equation}
\label{k_omega}
\begin{cases}
\omega = \omega_0 + \omega_1 \\
\bold k = \bold k_0 + \bold k_1
\end{cases}
\end{equation}
The wave 4-vector of the nonlinear wave $(\omega, \bold k)$ must also be isotropic,
which is required by the equality of the phase speed of the wave of the speed of light (1 in our units): $\omega^2 = \bold k^2$.
If $\bold k_1$ is parallel to $\bold k_0$, then as $\bold F$ we again obtain an ordinary transverse linear wave.
If the vectors $\bold k_1$ and $\bold k_0$ are not parallel, then $\bold k$ will no longer be transverse to $\bold A$.
Therefore, unlike the classical case of linear plane waves, a nonlinear plane wave (\ref{plosk_RUM}) may not be transverse.

Let's define a 4-current nonlinear wave (\ref{plosk_RUM}). According to (\ref{soput_tok}):
\begin{equation}
\label{tok_nelin}
\mathscr{J} = D \bold F = \big(-i (\bold F \bold k_1), i \omega_1 \bold F + \bold [\bold F \bold k_1] \big)
\end{equation}
From this it can be seen that the wave transverse criterion (\ref{plosk_RUM}) $\bold k_1 || \bold k_2$ is the absence of current: $\mathscr{J}=0$.

So, we see that nonlinear equations allow for the existence of longitudinal electromagnetic waves, but
only in the presence of regular currents.
The direction of propagation of a plane wave in such solutions may not coincide with the direction of energy transfer determined by the Umov-Poynting vector.

%%%%%%%%%% Twisting
\section{Twisting energy flow}\label{zakruch}

\textit{Twist} of the energy flow, in our terminology, is the rotor of Umov-Poynting vector $\bold S$, defined according to (\ref{Umov}):
\begin{equation}
\label{rotor_umov}
\bold \Omega = \nabla \times \bold S =\nabla \times [\bold E \bold H] = \big ( (\nabla \cdot \bold H) + \bold H \cdot \nabla\big) \bold E-
\big( (\nabla \cdot \bold E) + \bold E \cdot \nabla\big) \bold H
\end{equation}

Below we will show that for a certain, fairly wide class of free fields, the twisting of the energy flow is impossible for linear Maxwell equations, but possible for nonlinear ones. As such a class of fields we will consider flat normal fields.
\textit{Normal} field is free field whose (electric and magnetic) components are equal in magnitude and perpendicular to each other: $E=H, \bold E \perp \bold H$.
In the biquaternion representation, the normal field $\bold F$ is a null-vector field: $\bold F^2=0$. For important reasons, primarily the requirement of zero mass,
To describe light waves, it is normal fields that are required.
\textit{Flat} field we call a field that always remains in some fixed plane $\mathscr{P}$: $\bold E,\bold H \in \mathscr{P} \Leftrightarrow \bold F \in \mathscr{P }$. An example of linear plane fields is a classical plane electromagnetic waves. A nonlinear example of a normal plane field was considered in the previous section.

A flat normal field is generally represented as:
\begin{equation}
\begin{cases}
\label{plosk_norm}
\bold E = (\bold a cos{\phi} + \bold b sin{\phi}) f \\
\bold H = (\bold a sin{\phi} - \bold b cos{\phi}) f,
\end{cases}
\end{equation}
where $\bold a$ and $\bold b$ are two fixed real-valued mutually perpendicular unit vectors lying in the plane $\mathscr{P}$, and $f$ and
$\phi$ are some scalar differentiable functions $t$ and $\bold r$.
Another way (\ref{plosk_norm}) can be written as:
\begin{equation}
\label{plosk_norm2}
\bold F = (\bold a e^{i \phi} + \bold b e^{-i \phi}) f
\end{equation}
Based on the definition (\ref{rotor_umov}), we can obtain an expression for the field twist (\ref{plosk_norm}):
\begin{equation}
\label{rotor_umov2}
\bold \Omega = 2f (\bold b f_x - \bold a f_y),
\end{equation}
where $f_x = \frac{ \partial f}{\partial x}$ and $f_y = \frac{ \partial f}{\partial y}$.

It is not difficult to verify that if a normal plane field satisfies Maxwell linear equations (\ref{Maxwell}),
then it has no twist: $\bold \Omega=0$. At the same time, nonlinear equations (\ref{RUM}) admit solutions of the form (\ref{plosk_norm}), for which $\bold \Omega \ne 0$. Such solutions have the form of the
wave for which $f = f_1(t \pm z) f_2(x,y)$, which is easy to verify by directly substituting this solution into the equations.

The conclusion of this section is that in a number of important cases nonlinear equations, as opposed to linear ones,
allow solutions that have a twisting energy flow. Nonlinearity here manifests itself as the ability to form vortex structures of energy-momentum flow. Since this flow is largely determined by the Umov-Poynting vector (see Appendix \ref{Appendix_SP}) - we defined the twist in the form of a rotor of this vector.
The analysis presented here serves only to demonstrate, using specific examples, the fundamental difference between the solutions of nonlinear
equations from solutions of linear equations in terms of the possibility of the existence of vortex structures of a free field.

%%%%%%%%%%% central field %%%%%%%%%%%%%%%%%%%%%

\section{Central-symmetric field. Electromagnetic charge.}\label{centr}
We derived nonlinear equations (\ref{RUM10}) for a free field, i.e. field without singularities.
However, by virtue of their derivation, they also work in any 4-region lying outside the singularities
(although in the presence of singularities such a field can no longer be called free).
Now proceed with the search for a centrally symmetrical (in some reference frame)
solutions to these equations.
\begin{equation}
\label{centr_simm}
\bold F= f(t, r) \ \bold r
\end{equation}
where $f(t, r)$ is some complex-valued function differentiable with respect to both arguments.
As will be seen from what follows, at the central point $r=0$ such a field turns to infinity.
Therefore, we will look for a solution (\ref{centr_simm}) in the region outside the center-singularity.

For the field (\ref{centr_simm}) at $r \neq 0$ the following relations hold:
\begin{equation}
\begin{split}
& \bold g \equiv \nabla f = f_r \frac{\bold r}{r}\\
& \partial_t \bold F= f_t \bold r \\
& (\nabla \bold F) = 3 f + (\bold g \bold r) \\
& [\nabla \bold F] = [\bold g \bold r]
\end{split}
\end{equation}

When substituting (\ref{centr_simm}) into (\ref{RUM10}) after simple calculations we get:
\begin{equation}
\label{ur_f1}
\begin{cases}
Re\{  f^{\ast} f_t  \}=0\\
Re\{ f^{\ast}(3f + f_r r) \} = 0
\end{cases}
\end{equation}

The second of these equations is reduced to the form:
\begin{equation}
\label{ur_f}
Re\{ f^{\ast} f_r \} = - \frac{3 |f|^2}{r}
\end{equation}
Representing the function $f$ as:
\begin{equation}
f(t,r) = s e^{i\phi},\ s(t,r) \in \mathbb{R},\ \phi(t,r) \in \mathbb{R}
\end{equation}
Let's transform the equation (\ref{ur_f}) to the form:
\begin{equation}
Re\{ \frac{s_r}{s} + \frac{3}{r} \} = i \phi_r
\end{equation}
The last equation is satisfied only when both its sides are equal to 0, which follows:
\begin{equation}
\phi_r = 0 \ \Rightarrow \ \phi = \phi(t)
\end{equation}
\begin{equation}
  \frac{s_r}{s} = - \frac{3}{r} \ \Rightarrow \ s= \frac{\alpha}{r^3}, \ \alpha= \const \in \mathbb{R}
\end{equation}
\begin{equation}
\label{f_ur}
f = \frac{\alpha}{r^3} e^{i \phi(t)}
\end{equation}
It is easy to check that for $f$ in (\ref{f_ur}) the first of the equations (\ref{ur_f1}) is always satisfied as well.

Thus, the general centrally symmetric solution (\ref{RUM10}) has the form of an ordinary Coulomb field with a phase factor $e^{i \phi(t)}$:
\begin{equation}
\label{kulon1}
\bold F = \frac{\alpha \bold r}{r^3} e^{i \phi(t)}
\end{equation}
Since we found the solution (\ref{kulon1}) in a certain frame of reference, it is necessary to give its Lorentz-covariant form.
This is:
\begin{equation}
\label{kulon2}
\bold F = \bold F_0 e^{i \Phi}
\end{equation}
where $\bold F_0$ is the usual Coulomb field transformed to a given reference frame according to (\ref{PL_pole}),
and $\Phi$ is some Lorentz invariant.
If we assume that the singularity at the center of the field (\ref{kulon2}) is associated with some real 4-vector $K=(\omega, \bold k)$,
then the Lorentz invariant has the form: $\Phi = \omega t \pm \bold k \cdot \bold r$, and \textit{nonlinear Coulomb field}
(\ref{kulon2}) has a wave character:
\begin{equation}
\label{kulon3}
\bold F = \bold F_0 e^{i ( \omega t \pm \bold k \cdot \bold r)}
\end{equation}
It is noteworthy that this is a solution associated the usual Coulomb field in the sense we defined above in (\ref{nelin_resh}).

The resulting field is associated with some singularity, i.e. with some charged particle.
Determine how the 4-momentum of this Coulomb particle $P$ and the wave 4-vector $K$ relate to each other.
In the particle's rest frame of reference, all directions are the same. Therefore, it contains $\bold k_1=0$, and
the field has the form: $\bold F = \bold F_0 e^{i \omega t} = \frac{\alpha \bold r}{r^3} e^{i \omega t}$
When transitioning to a reference frame moving with speed $\bold V$, $\bold F_0$ is transformed into a Coulomb field
charge moving at a speed of $\bold v = -\bold V$. The 4-vector $K$ is then transformed so that $\bold k = \omega \bold v$.
The latter means:
\begin{equation}
\label{kulon4}
K = \lambda P
\end{equation}
where $\lambda$ is a real-valued constant defined for a given particle.

The nonlinear Coulomb field (\ref{kulon1}) can be interpreted as the Coulomb field of the complex charge $q$:
\begin{equation}
\bold F = \frac{q \bold r}{r^3},
\end{equation}
The complex-valued charge $q$ has the meaning of an electromagnetic charge, variably combining electric and magnetic charges:
\begin{equation}
q = \alpha e^{i \phi(t)}
\end{equation}
With a linearly varying phase, the complex charge periodically passes through all complex values of a given modulus.
For real values, $q$ becomes an electric charge, and for imaginary values, a magnetic charge.
However, as it turns out, when interacting with other charges, it is not the phase of the charge itself that is important, but how the phases of different charges relate to each other.
As will be shown below (Section 14), the simplest matching of the fields of two charges leads to matching their phases, resulting in effectively electrical interaction between them.

In the next section, using the appropriate generalized functions, we will extend the resulting solution outside the singularity to the singularity itself.

%%%%%%%%%%% Equations with one singularity
\section{Equations with one singularity.}\label{Singular}
In section \ref{Free} we derived the field equations in "vacuum", i.e. in a space-time region that does not contain field singularities.
Let's assume now that in the region under consideration there is one point (centrally symmetric) singularity.
In the presence of a singularity, the field energy in the region containing this singularity is no longer conserved. The following visual model of energy exchange between two types, or \textit{planes} of movement: field and kinematic is useful.
The field singularity plays the role of a channel through which energy can flow from the field \textit{plane} to the kinematic \textit{plane} and back. Kinematically
the singularity of the field is represented by a particle\footnote{Hereinafter, a particle is understood as a point-like charged particle.}.
We assume, and this is largely confirmed by experience, that apart from singularities there are no other channels of energy exchange between the field and kinematic \textit{planes}.
Therefore, energy cannot flow out of the field through regular currents.

As we saw in the previous section, outside the point singularity the field is Coulomb with a phase factor (\ref{kulon3}).
As a four-dimensional region containing the singularity line, we take a certain hypercylinder $V_4$ of "height" \ (in a certain reference frame) $\Delta t = t_2 - t_1$, limited "above and below" \ by two time slices, i.e. spatial volumes $S^{(1)}_3$ and $S^{(2)}_3$ (Fig.\ref{Ris1}).
  \begin{figure}
\centering
\includegraphics[width=2.755in,height=2.32in]{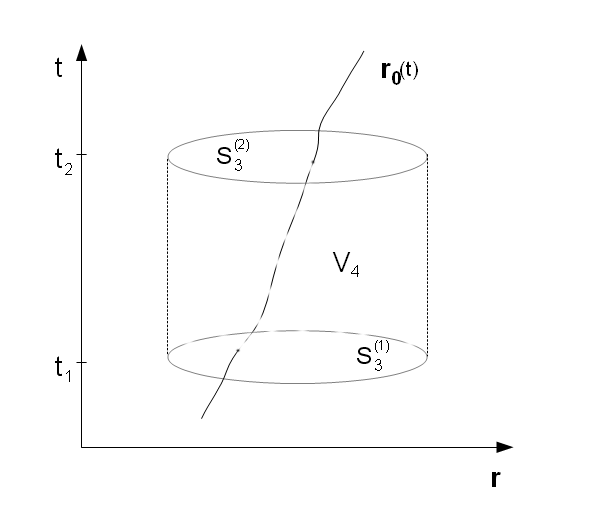}\caption{Region with singularity}\label{Ris1}
\end{figure}

The rate of energy-momentum flow from a field to a particle or from a particle to a field in a 4-volume $V_4$ is defined as $\frac{d P}{d t} \Delta t$,
where $P=(\epsilon, \bold p)$ is the 4-momentum of the particle.
For the energy balance, instead of (\ref{sigma_int}) we get:
\begin{equation}
\label{sigma_int_zar}
\frac{1}{2} \oint_{S_3} \bold F^{\ast} dZ \bold F \ = \ \frac{d P}{d t} \Delta t
\end{equation}
where $S_3$ is the complete hypersurface bounding $V_4$.

The Ostrogradsky-Gauss biquaternion formula (\ref{OG}), formulated for continuously differentiable functions,
can be transferred to the case of singular functions using the apparatus of generalized functions. Considering the function $\bold F(Z)$ in (\ref{sigma_int_zar}) as a generalized one, we can
rewrite the integral over the hypersurface $S_3$ as an integral over the 4-volume, as we did above in (\ref{sigma_int2}) for a regular function:
\begin{equation}
\label{sigma_int_zar3}
  \oint_{S_3} \bold F^{\ast} dZ \bold F = \int_{V_4} ( \bold F^{\ast} D \bold F) dV_4
\end{equation}

Assuming that the time interval $\Delta t$ is small enough to neglect the change in $(\bold F^{\ast} dZ \bold F)$ on it, the integral (\ref{sigma_int_zar3}) can be expressed as the product integral over spatial volume $V_3 = S^{(1)}_3 = S^{(2)}_3$ for time:
\begin{equation}
\label{sigma_int_zar4}
\int_{V_4} ( \bold F^{\ast} D \bold F) dV_4 \ = \Delta t \int_{V_3} ( \bold F^{\ast} D \bold F) dV_3
\end{equation}
Then (\ref{sigma_int_zar}) is transformed to the form:
\begin{equation}
\label{delta0}
\int_{V_3} ( \bold F^{\ast} D \bold F) dV_3 \ = \ \frac{d P}{d t}
\end{equation}

In a given 4-domain (not containing other singularities), at all points outside the trajectory of the singularity $\bold r_1(t)$ the function $(\bold F^{\ast} D \bold F)$ turns to 0. This allows us to replace the indeterminate at the center a regular function $(\bold F^{\ast} D \bold F)$ to a generalized $\delta$-function:
\begin{equation}
\label{delta1}
(\bold F^{\ast} D \bold F) \ = \ A(t) \ \delta (\bold r - \bold r_1 (t))
\end{equation}
where $A(t)$ is some regular function of time. Substituting (\ref{delta1}) into (\ref{delta0}), we find this function: $A(t)= \frac{d P}{d t}$, and as a result from (\ref{delta0}) we get :
\begin{equation}
\label{delta2}
(\bold F^{\ast} D \bold F) \ = \ \frac{d P}{d t} \delta (\bold r - \bold r_1 (t))
\end{equation}
This is \textit{Maxwell nonlinear equation with the right side}, which serves as an analogue of the equation (\ref{um1}) in the presence of a singularity.

Note, in the equation (\ref{delta2}), the field $\bold F$ is united or common - there is no division into the particle field and the external field.
Such a conditional division can be carried out in the \textit{quasilinear case}, when
the common field can be represented as the sum of the fields of the particle $\bold F_1$ and the external field $\bold F_{e}$:
\begin{equation}
\label{q_lin}
\bold F = \bold F_1 + \bold F_{e}
\end{equation}
so that each of them separately satisfies the nonlinear equation in emptiness (\ref{um1}),
\begin{equation}
\label{q_lin2}
\begin{cases}
(\bold F_1^{\ast} D \bold F_1) = 0 \\
(\bold F_{e}^{\ast} D \bold F_{e}) = 0
\end{cases}
\end{equation}
and the total field $\bold F$ satisfies a nonlinear equation with singularity (\ref{delta2}).

Note that in order to obtain the charge's own field $\bold F_1$, satisfying the first of the equations (\ref{q_lin2}), it is sufficient to supplement the centrally symmetric solution (\ref{kulon3})
zero value at the center of the singularity: $\bold F_1(t,\bold r_1(t))= 0$. The last condition means absence of self-interaction.
Then, energy exchange through the singularity, described by the equation (\ref{delta0}), will be reduced to an equation containing only
\textit{field interaction}:
\begin{equation}
\label{delta3}
  \frac{1}{2} (\bold F_1^{\ast} D \bold F_{e}) + (\bold F_{e}^{\ast} D \bold F_1) \ = \ \frac{d P}{d t} \delta (\bold r - \bold r_1 (t))
\end{equation}
or
\begin{equation}
\label{delta4}
Re\{ \bold F_{e}^{\ast} \mathscr{J}_1 + \bold F_1^{\ast} \mathscr{J}_{e} \} \ = \ \frac{d P}{ d t} \delta (\bold r - \bold r_1 (t))
\end{equation}
where $\mathscr{J}_1 = D \bold F_1$ is the particle current (which for nonlinear solutions contains both singular and regular parts), $\mathscr{J}_{e} =D \bold F_{ e}$ is the regular external field current.
Each of the terms of the form $\bold F^{\ast}\mathscr{J}$ in (\ref{delta4}) is an expression of the work (per unit time) of the corresponding field applied to the "counterpart" \ current.

One can represent the current $\mathscr{J}_1$ as the sum of singular and regular currents:
\begin{equation}
\label{delta411}
\mathscr{J}_1 = \mathscr{J}_{1sng} + \mathscr{J}_{1reg}
\end{equation}
and according to this, we divide the equation (\ref{delta4}) into singular and regular parts:
\begin{equation}
\label{delta41}
\begin{cases}
Re\{\bold F_{e}^{\ast} \mathscr{J}_{1sng} \} \ = \ \frac{d P}{d t} \delta (\bold r - \bold r_1 (t) )\\
Re\{\bold F_{e}^{\ast} \mathscr{J}_{1reg} + \bold F_1^{\ast} \mathscr{J}_{e} \} \ = \ 0
\end{cases}
\end{equation}
The meaning of the first of the equations (\ref{delta41}) is ordinary: it expresses the fact that the transfer of energy to a charged particle is equal to the work of the external field on the charge.
The second of these equations has a purely nonlinear nature and means mutual compensation of the work of the particle field and the external field on
"counterpart" \ regular current.

We can see how the equations (\ref{delta41}) in the special case yield the usual equations of motion of a charged particle in an external field.
To do this, we need to take a linear external field: $\mathscr{J}_{e} = 0$, and a purely electric current of the singularity: $Im \mathscr{J}_1 = 0$.
Under these conditions, the second equation (\ref{delta41}) is satisfied trivially, and from
the first equation (\ref{delta41}) and Lorentz covariance considerations for the 4-singularity current we obtain:

\begin{equation}
\label{tok_polya_1}
\mathscr{J}_{1sng} =  \   e \gamma (1,\bold v) \delta (\bold r - \bold r_1 (t))
\end{equation}
The first of the equations (\ref{delta41}) takes on the known form of the Lorentz force and its work per unit time on an electric charge of magnitude $e$ from the external field $\bold F_e = \bold E_{e} + i \bold H_{e }$:
\begin{equation}
\label{delta5}
\begin{cases}
e \bold v \cdot \bold E_e \ = \ \frac{d \epsilon}{d t}\\
e \bold E_e + e \bold v \times \bold H_e \ = \ \frac{d \bold p}{d t}
\end{cases}
\end{equation}
where $\bold v = \frac{d \bold r_1}{d t}$ is the velocity of the singularity, $\gamma = 1/\sqrt{1 - v^2}$, and $e$ is some real-valued constant.

Recall that the singularity field has a centrally symmetric form (\ref{kulon1}).
From the identity\footnote{This identity is proven similarly to the more famous identity $\Delta \frac{1}{r} = - 4 \pi \delta(\bold r)$}
\begin{equation}
\nabla \cdot \frac{ \bold r}{r^3} = 4 \pi \delta(\bold r)
\end{equation}
it follows that the current for the field (\ref{kulon1}), extended to the central point, in the rest frame is equal to:
\begin{equation}
  \mathscr{J}_{1sng} = D \bold F = \nabla \cdot \frac{ \alpha \bold r}{r^3} = 4 \pi \alpha \delta(\bold r)
\end{equation}
Comparing the latter with (\ref{tok_polya_1}), we get:
\begin{equation}
e = 4 \pi \alpha
\end{equation}
The constant $\alpha$ in (\ref{kulon1}) coincides with the charge $e$ in (\ref{tok_polya_1})
  (up to a factor of $4 \pi$, which is due to our choice of system of units).
This means that the magnitude of the particle charge as a field source
coincides with the value of its "test" \ charge in the external field.

We need to make sure that the velocity of the singularity $\bold v$ coincides with the velocity of the particle $\frac{\bold p}{\epsilon}$.
From the constancy of the particle mass $\epsilon^2 - {\bold p}^2 = m^2$ the following relation follows:
\begin{equation}
\label{delta6}
\epsilon \frac{d \epsilon}{d t} = \bold p \cdot \frac{d \bold p}{d t}
\end{equation}
Multiplying the first of the equations (\ref{delta5}) by $\epsilon$, and the second scalar by $\bold p$ and applying (\ref{delta6}), we arrive at the following
equation:
\begin{equation}
\label{delta7}
\bold E_e \cdot (\epsilon \bold v - \bold p) = \bold p \cdot [\bold v \bold H_e]
\end{equation}
This relation must hold for any external field $\bold E_e, \bold H_e$, which is only possible for $\bold v = \frac{\bold p}{\epsilon}$.

So, we are convinced that in the linear approximation the transition from nonlinear equations to ordinary equations for charge in an external field is carried out.
With this transition, instead of one nonlinear equation, one comes up with two separate equations:
the classical Maxwell equation with the right-hand side (which in our case is the definition of current) and
energy balance equation or Lorentz force.

%%%%%%%% Matched fields.
\section{Coherent fields. Electromagnetic charge}\label{Agreed}

We can apply the same reasoning that we used in the previous section when deriving the formula (\ref{delta2}) to the case when in the region under consideration there are $n$ singularities described by trajectories $\bold r_i(t)$. As a result, we obtain the Lorentz-covariant equation:
\begin{equation}
\label{delta_para}
  \frac{1}{2} (\bold F^{\ast} D \bold F) \ = \ \sum\limits_{i=1}^n \frac{d P_i}{d t} \delta (\bold r-\bold r_i(t))
\end{equation}
where $P_i$ are the 4-momenta of each of the particles corresponding to these singularities.

Now we consider the quasilinear case of two \textit{identical particles}. Wherein
we will proceed from the assumption that the general field $\bold F$ can be obtained as the sum of the fields $\bold F_1$ and $\bold F_2$ from each singularity separately:
\begin{equation}
\label{sogl1}
\bold F = \bold F_1 + \bold F_2
\end{equation}
Let us consider some 4-region lying \textit{outside the trajectories of both singularities}. As the individual fields of each of the singularities, we take their wave Coulomb fields (\ref{kulon3}):
\begin{equation}
\label{sogl2}
\begin{cases}
\bold F_1 = f_1 \bold F_{1_0}, \ f_1 = e^{i \phi_1} = \bold F_{1_0} e^{i ( \omega_1 t \pm \bold k_1 \cdot \bold r)} \ \
\bold F_2 = f_2 \bold F_{2_0}, \ f_2 = e^{i \phi_2} = \bold F_{2_0} e^{i ( \omega_2 t \pm \bold k_2 \cdot \bold r)}
\end{cases}
\end{equation}
where $\bold F_{1_0}$ and $\bold F_{2_0}$ are the usual Coulomb fields of each particle in the reference frame under consideration.
The plus or minus sign within the wave phases for each of the fields will be determined below.

By analogy with the second equation (\ref{delta41}), the nonlinear interaction of fields (outside singularities) will be written as:
\begin{equation}
\label{sogl3}
Re\{\bold F_1^{\ast} \mathscr{J}_2 + \bold F_2^{\ast} \mathscr{J}_1 \} \ = 0
\end{equation}
where $\mathscr{J}_1 = D \bold F_1$ and $\mathscr{J}_2 = D \bold F_2$ are the regular currents of each particle. Based on (\ref{soput_tok}) and (\ref{sogl2}) for the current $\mathscr{J}_1$ we obtain:
\begin{equation}
\label{sogl4}
\mathscr{J}_1 = i f_1 \bold F_{1_0} \overline K_1
\end{equation}
where $K_1= K_1(t) = (\omega_1, \bold k_1)$.
A similar expression is obtained for the current $\mathscr{J}_2$.
The equation (\ref{sogl3}) is rewritten as:
\begin{equation}
\label{sogli5}
Im\{ f A + f^{\ast} B \} \ = 0
\end{equation}
where $f=f^{}_2 f^{\ast}_1$ and
\begin{equation}
\label{sogl6}
\begin{cases}
A = \bold F_{2_0} \overline K_2 \bold F^{\ast}_{1_0} \\
B = \bold F_{1_0} \overline K_1 \bold F^{\ast}_{2_0}
\end{cases}
\end{equation}
Denote $ \Delta K = K_2 - K_1$. Then $A$ can be represented as:
\begin{equation}
\label{sogli7}
A = \bold F_{2_0}( \overline K_1 +\overline {\Delta K} ) \bold F^{\ast}_{1_0} = B^{\ast} + \bold F_{2_0} \overline { \Delta K} \bold F^{\ast}_{1_0}
\end{equation}
The result (14.6) will be expressed as:
\begin{equation}
\label{sogl8}
Im\{ f B^{\ast} + f^{\ast} B + f \bold F_{2_0} \overline {\Delta K} \bold F^{\ast}_{1_0} \} \ = 0
\end{equation}
or
\begin{equation}
\label{sogl88}
Im\{ f \bold F_{2_0} \overline {\Delta K} \bold F^{\ast}_{1_0} \} \ = 0
\end{equation}

Consider the movement of particles in the system of their center of mass.
Since the particles are assumed to be identical, then the 4-momentum and 4-wave vector of each
of them are related by the equation (\ref{kulon4}) to the same $\lambda$. Hence,
$\bold k_1 = -\bold k_2 = \bold k$, $\omega_1 = \omega_2$, $\Delta K = -2 \bold k$.
\begin{equation}
\label{sogl89}
  f \bold F_{2_0} \overline {\Delta K} \bold F^{\ast}_{1_0} \ =-2 f \big ( i [\bold F^{\ast}_{1_0} \bold F_{2_0}] \cdot \bold k,
( \bold F_{2_0} \bold k) \bold F^{\ast}_{1_0} + \bold F_{2_0} ( \bold F^{\ast}_{1_0} \bold k) - \bold k (\bold F^{\ast}_{1_0} \bold F_{2_0} ) \big )
\end{equation}

  \begin{figure}
\centering
\includegraphics[width=2.755in,height=1.807in]{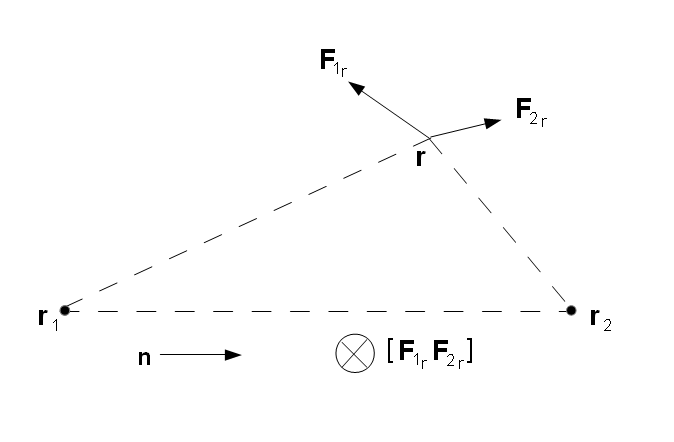}\caption{Two charges}\label{Ris2}
\end{figure}
The fields of moving charges in the center of mass system $\bold F_{1_0}, \bold F_{2_0}$ are obtained from the fields of stationary charges $\bold F_{1_r}, \bold F_{2_r}$
using two suitable boosts of opposite speeds $\bold V$ and $-\bold V$ (\ref{PL_pole}):
\begin{equation}
\label{sogl10}
\begin{cases}
\bold F_{1_0} = \ch 2 \theta \bold F_{1_r} - 2 \sh^2 \theta ( \bold F_{1_r} \bold n) \bold n + i \sh 2 \theta [ \bold F_{1_r} \bold n ] \\
\bold F_{2_0} = \ch 2 \theta \bold F_{2_r} - 2 \sh^2 \theta ( \bold F_{2_r} \bold n) \bold n - i \sh2 \theta [ \bold F_{2_r} \bold n ]
\end{cases}
\end{equation}
where the boost parameters $\bold n$ and $\theta$ are related to its speed by the relation $\bold V = \bold n \th 2\theta$. Obviously, $\bold k \parallel \bold n$.

$\bold F_{1_r}$ and $\bold F_{2_r}$ are the usual (real-valued) Coulomb fields of each particle, each taken in the rest frame of its particle, in which they are directed along their radius vectors.
The latter can be expressed by using the Lorentz transformation (\ref{PL_koord}) through the coordinates of the point under consideration $(t,\bold r)$.
Thus, the relative radius vector $\bold R'_1$ for $\bold F_{1_r} = \frac{ \alpha \bold R'_1 } { {R'_1}^3}$ is equal to
\begin{equation}
\bold R'_1 = \bold r' - \bold r'_1 = \bold R_1 - \bold n ( t \sh 2 \theta - 2 (\bold R_1 \bold n) \sh^2 \theta )
\end{equation}
where $\bold R_1 = \bold r - \bold r_1$ is the radius vector in the center of mass system.
We have a similar expression for $\bold R'_2$ - the radius vector defining $\bold F_{2_r}$. Important,
that the vectors $\bold F_{1_r}$ and $\bold F_{2_r}$ lie in the plane of the vectors $\bold r$ and $\bold n$ shown in Fig.\ref{Ris2}.
This is quite obvious from the fact that these vectors are directed along the corresponding radius vectors $\bold R'_1$ and $\bold R'_2$
in their rest frames, and the radius vector, when transformed into a center of mass reference frame, does not acquire
component transverse to $\bold n$.
From this, according to (\ref{sogl10}), it follows:
\begin{equation}
\begin{cases}
[\bold F^{\ast}_{1_0} \bold F_{2_0}] \cdot \bold k = 0\\
Im\{ (\bold F^{\ast}_{1_0} \bold F_{2_0} ) \} = 0
\end{cases}
\end{equation}
Then, taking into account (\ref{sogl89}), condition (\ref{sogl88}) reduces to the reality of the function $f$:
\begin{equation}
\label{soglas20}
Im f = 0
\end{equation}
Because
\begin{equation}
f = e^{i( \phi_2(t,\bold r) - \phi_1(t,\bold r)) }
\end{equation}
then the condition (\ref{soglas20}) means \textit{phase matching} of the wave fields of singularities for any values of $t$ and $\bold r$:
\begin{equation}
\label{soglas201}
\phi_1(t,\bold r) = \phi_2(t,\bold r)
\end{equation}
As can be seen from what follows, an additional non-zero phase shift $2 \pi n$ is impossible.

Now we can turn to the question of which sign ("plus" \ or "minus") should be chosen for the phases of each of the particles in their wave functions
(\ref{sogl2}). Let's take the minus sign for the first particle:
\begin{equation}
\label{sogl21}
\phi_1 = \omega_1 t - \bold k_1 \cdot \bold r
\end{equation}
From here, from (\ref{soglas201}) in the center of mass system for the second particle we obtain:
\begin{equation}
\label{sogl22}
\phi_2 = \phi_1 = \omega_2 t + \bold k_2 \cdot \bold r
\end{equation}
Then in its phase the plus sign must be selected.
We come to the important conclusion that in order to coordinate the fields of particles, their wave phases must be \textit{conjugate}:
the minus sign of the first particle corresponds to the plus sign of the second, and vice versa -
the plus sign for the first particle corresponds to the minus sign for the second particle.
The first of these options is written as:
%%%
\begin{equation}
\label{sogli79}
\begin{cases}
\bold F_1 = \bold F_{1_0} e^{i ( \omega_1 t - \bold k_1 \cdot \bold r)} \\
\bold F_2 = \bold F_{2_0} e^{i ( \omega_2 t + \bold k_2 \cdot \bold r)}
\end{cases}
\end{equation}
%%%

From the above analysis it follows that phase matching or \textit{coherence} of fields is required for a quasi-linear representation of the total field in the form of a sum of the fields of each particle (\ref{sogl1}). If the fields are not coherent, then such a representation is impossible, which effectively requires
introduction of a third field in total. The latter field should be the field of radiation. The work to compensate for this field can be assessed as the left side (\ref{sogl3}), which therefore plays the role of nonlinear interaction of incoherrent  fields. Based on the above, it can be assumed that
Coherrent fields minimize nonlinear interaction, while fields incoherrence leads to radiation.

The minus and plus signs within the wave phase obviously describe waves traveling in the direction of the particle's velocity and in the opposite direction, respectively.
When the phases of particles moving towards each other are coordinated, their waves turn out to be traveling in the same direction.

Above we studied the joint field of two identical particles in the quasilinear case in the region outside singularities. Let's see what happens now
on the singularities themselves. From the first equation (\ref{delta41}) applied to the first of the singularities, we get
the following equation of motion of its particle:
\begin{equation}
Re\{\bold F_{2}^{\ast} \mathscr{J}_{1sng} \} \ = \ \frac{d P_1}{d t} \delta (\bold r - \bold r_1 (t) )
\end{equation}
But since the phases $\mathscr{J}_{1sng}$ and $\bold F_{2}$ for matched fields cancel each other,
we arrive at the equations of motion of one charged particle under the influence of the electric Coulomb field of another charged particle,
which take place in ordinary electrodynamics.

Quasilinear field matching "hides" \ nonlinear effects in observing the motion of charged particles,
and instead of interaction of electromagnetic charges, in this case we obtain the interaction of ordinary electric charges,
as stated above at the end of Section \ref{centr}.
However more complex nonlinear cases of interaction of charges are possible, in which their electromagnetic nature can manifest itself.
In this work we do not go with such cases, leaving their study for the future.

%%

%%%%%%%%%%%% Conclusions
%\section{Conclusions}\label{Vyvody}
\hfill

\bgroup\obeylines
\textbf{
Conclusions
}
\textbf{

}
\egroup

The main original goal of this work was to show that Maxwell equations are a consequence of
the principle of conservation of field energy flow. However, by solving this problem, we obtained broader nonlinear equations.
In general form, including equations (\ref{um1}) for a free field
and equations (\ref{delta_para}) for a field with $n$ singularities, these equations have the form:
\begin{equation}
\label{RUM_res}
\frac{1}{2} (\bold F^{\ast} D \bold F) \ = \ \sum\limits_{i=1}^n \frac{d P_i}{d t} \delta (\bold r-\bold r_i(t)),\ n \geqslant 0
\end{equation}
It is important that in these equations $\bold F$ there is a single field, not divided into free fields and fields of individual particles.

Maxwell classical linear equations turn out to be embedded in these nonlinear equations in a special way and are their limiting case.
The physical meaning of the free field equations we obtained is that this field can be created by its own source structures,
on which it does not perform work. We defined this relationship between the field and its own sources as recursive.

It is well known that light is an essentially spatiotemporal phenomenon in which it is impossible to
separate spatial and temporal components from each other. But in the classical Maxwell equations for a field in emptiness (\ref{Maxwell}), the differentiation operators with respect to the spatial coordinate and time are linearly separated
(come additively). At the same time, corresponding nonlinear equations (\ref{RUM}) are characterized by the inseparability of these operators, since they enter the equations in a more complex form than a simple sum. This suggests that nonlinear equations are more appropriate than linear equations for describing real light.

We called field own sources regular charges and currents - in contrast to ordinary singular charges and currents.
Classical electrodynamics deals
either with singular sources or with a complete absence of sources. Our study examines
the theoretical possibility of the existence of non-singular (regular) field sources.

Nonlinear free field equations admit among their solutions plane waves having a longitudinal component, i.e. which ones are not completely transverse.
The criterion for the existence of such solutions is the presence of a non-zero regular current. Otherwise, we have an ordinary transverse wave.
The presence of a longitudinal component in light waves can explain in classic way the fact that light has its own angular momentum,
which is impossible in the case of purely transverse waves.

In section \ref{zakruch} it was found that if we limit the consideration to flat normal fields, then nonlinear equations, unlike linear ones
allow solutions that have a twisting energy flow. Nonlinearity here manifests itself as the ability to form vortex structures of energy-momentum flow, which is largely determined by the Umov-Poynting vector - we define the twist in the form of a rotor of this vector.
Specific examples show the fundamental difference between solutions of nonlinear
equations from solutions of linear equations in terms of the possibility of the existence of vortex structures of a free field.

Particular attention in the work is paid to the problem of dividing the field into the "intrinsic" \ field of a charged particle
and a field "external" \ to it.
From the nonlinear field equations follow both the classical Maxwell equations themselves and the equations of charge moving under
the action of Lorentz force.
Thus, we have given a solution to the problem posed in the introduction - to find nonlinear field equations that include interaction.
Unlike in classical electrodynamics, our equations clearly show why the "test" charge of a particle and its charge as a field source coincide.
Another important consequence of the inclusion of interaction in the fundamental equations of field is unnecessity to use Lagrangian formalism.

The simplest, nevertheless important, class of nonlinear fields are associated solutions of the form (\ref{nelin_resh}).
These are ordinary Maxwell fields endowed with a wave factor.
The existence of such solutions shows, at least for the case of two particles we studied, that the known global dual transformations have a generalization in the form of local phase transformations.

The appearance of field solutions of the form of a Coulomb wave (\ref{kulon3}), combining the field of a point charge and a wave,
brings together classical electrodynamics and quantum mechanics and indicates the possibility of an electromagnetic description of
de Broglie wave of the electron. This allows us to talk about a different view of wave-particle duality.
The concept of an electromagnetic wave in nonlinear electrodynamics is thus
is not limited to light, but also applies to particle fields. Moreover, the waves describing these two fundamental types of field structures
fundamentally different.

In the works \cite{Bial},\cite{Raymer} it was shown that the field of electromagnetic wave itself, in a certain sense, plays a role
quantum mechanical wave function of a photon: the square of its modulus determines the mathematical expectation of the energy density at a given point in space.
The nonlinear solution we obtained (\ref{kulon3}) indicates that for a charged particle, a similar "energy" \ wave function can be specified by its complex-valued electromagnetic field.

Table \ref{table1} provides a summary of the main differences of nonlinear electrodynamics resulting from
equations (\ref{RUM_res}), from classical linear electrodynamics\footnote{The curl or twist of the energy flow is understood as the rotor of the Umov-Poynting vector, and possible solutions are limited to the class of normal plane fields.}.
\begin{table}[h!]
   \begin{center}
     \caption{}
     \label{table1}
     \begin{tabular}{l|l|l} % <-- Alignments: 1st column left, 2nd middle and 3rd right, with vertical lines in between
        & Linear ED & Nonlinear ED\\
       \hline
Basic principles & eq. Maxwell + interaction & conservation of energy\\
Energy representation & no & exists\\
Space and time operators & separate & inseparable\\
Description of particles and fields & separate & united \\
Light & transverse waves & + longitudinal waves \\
Twisting energy flow & no & exists \\
Charged particle & point singularity & + wave \\
Charges and currents & singular & + regular \\
Charges & electric & electromagnetic \\
Charge interaction & Coulomb & + phase
     \end{tabular}
   \end{center}
\end{table}

In addition to describing a particle as an electromagnetic wave, nonlinear electrodynamics based on equation (\ref{RUM_res}) exhibits a number of other similarities with quantum field theory.
Thus, phase coherence of two identical charged particles
is a classical analogue of the Pauli principle: the fields of these particles
are consistent when their phases are conjugate.

Phase matching leads to the existence, along with the usual Coulomb interaction, of phase interaction,
similar to quantum spin interaction (also called "exchange" interaction).
In both cases, the interaction is determined by the overlap of the corresponding wave functions.
The sign "$\pm$" \ inside the wave phase expresses a discrete degree of freedom,
similar to spin 1/2.

The complexity of the charge and the presence of its phase is perhaps the most striking consequence of the nonlinearity of the field equations we obtained.
In our approach, the charge of the particle is electromagnetic, periodically passing through various linear combinations of electric and magnetic charges from purely electric to purely magnetic.
In real processes of interaction of charged particles with each other and the field, it is not the charge of the particle itself that plays a role, but its phase relationship with other charges and fields. In this way, the old riddle of the "absence" \ of magnetic charges in nature (while they are present in theory) is resolved.

Within our approach, there are neither purely electric nor purely magnetic charges. The usual interaction of charged particles, due to the requirements of energy efficiency, is reduced to an interaction similar to the Coulomb interaction of electric charges. In other words, electromagnetic charges "in ordinary life" \ look like electric ones. With all this, we can expect the observation of new physical effects of nonlinear interaction, in which the electromagnetic nature of the charges will manifest itself - an interaction in which the phases of the particles do not have time to adjust to each other.

According to quantum field theory, strong electromagnetic fields polarize the vacuum, which in turn
leads to the appearance of corrections in the Maxwell Lagrangian of these fields \cite{Grib}.
Taking into account the effects of vacuum polarization can be interpreted as nonlinearity of electrodynamics.
"Quantum effects create a non-zero right-hand side (current) even in the absence of charged particles in the initial state" \cite{Zeld}.
This allows us to correlate our regular currents with vacuum currents.
Both types of current have an additional effect compared to ordinary "external" \ currents.

Anyhow, there is a fundamental difference between these two approaches.
In our case, continuous currents exist initially in Maxwell nonlinear equations, while
in quantum field theory, Maxwell linear equations are supplemented by currents built on the summation of the set
discrete events of the birth of particle-antiparticle pairs.

In this article, we studied singularities of the point-centric symmetric type, which describe point-charge  particles.
For complete picture it requires the study of an axis-centric symmetric solution of nonlinear equations which would describe magnetic moment of the particle. In the same scope lie questions about the connection between the nonlinear equations we obtained and the angular momentum of the electromagnetic field including its spin.
Also, a perspective direction of research should become a nonlinear description of radiation processes.
Discovering these and other questions will help reveal deeper connections between classical nonlinear electrodynamics and quantum field theory
and take full advantage of the capabilities of Maxwell equations and their generalizations.

The author thanks A.V. Goryunov for discussion and useful advice.

%%%%%%%%%%%%%%%%%%%%%%%%%%%%%
%%%%%%% APPLICATIONS %%%%%%%%%%%%%%
%%%%%%%%%%%%%%%%%%%%%%%%%%%%%
%\section{Applications}

\newpage

\hfill

\bgroup\obeylines
\textbf{
Appendices
}
\textbf{

}
\egroup

%%%%%%%%%%%% Appendix 1 - biquaternion algebra
\subsection{Algebra of biquaternions}\label{Appendix_biq}

As in our previous work \cite{Kot} (p. 160) we use scalar-vector
representation of biquaternions, 
in which the biquaternion $\mathscr{B}$ is a bundle of complex
number $s$
and three-dimensional complex vector $\bold{u}$:

\begin{equation}
\label{refText2}
\mathscr{B} = \ (s, \bold{u})\ {\equiv}\ s + \bold{u},
\ s \in \mathbb{C}, \bold{u} \in  \mathbb{C}^3
\end{equation}
The product of two biquaternions $\mathscr{B}_1 = (s_1, \bold{u}_1)$ and
$\mathscr{B}_2 = (s_2, \bold{u}_2)$ is defined as:
\begin{equation}
\label{refText3}
\mathscr{B}_1 \mathscr{B}_2\
\ =
\ (s_1 s_2 +( \bold{u}_1 \cdot \bold{u}_2),s_1
 \bold{u}_2+s_2 \bold{u}_2 + i( \bold{u}_1 \times \bold{u}_2)),
\end{equation}
where $( \bold{u}_1 \cdot \bold{u}_2)$ and
$(\bold{u}_1 \times \bold{u}_2)$ - scalar and vector
the products of $ \bold{u}_1$ and $ \bold{u}_2$ respectively.

In this representation, quaternions, as a special case of biquaternions, have the form:
\begin{equation}
\label{quater_def}
Q= (\alpha, i \bold a), \alpha \in \mathbb{R}, \bold{a} \in \mathbb{R}^3
\end{equation}
It is important to distinguish quaternions from real biquaternions, which have the form:
\begin{equation}
\mathscr{B} = (\alpha, \bold a), \ \alpha \in \mathbb{R}, \bold{a} \in \mathbb{R}^3
\end{equation}
Real biquaternions play a special role in our approach since the space-time coordinate
$Z$ and energy-momentum $K$ are just such quantities.

The space of biquaternions has a basis \cite{Imaeda}(p.141) $e_n$, $n=0,1,2,3$: $e_0=1$ and for $k>0$:
\begin{equation}
\label{basis}
\begin{cases}
e_k^2 = 1\\
e_i e_j = - e_j e_i = i e_k,
\end{cases}
\end{equation}
where $i,j,k$ are cyclically permutable indices 1,2,3. The basis biquaternions $e_n$ are real. An arbitrary biquaternion $\mathscr{B}$ is expanded over this basis:
\begin{equation}
  \mathscr{B} = \sum\limits_{n=0}^3 e_n b_n,\ b_n \in \mathbb{C}
\end{equation}
Thus, for a real-valued space-time coordinate $Z$ we have:
\begin{equation}
Z = (t,\bold r) = \sum\limits_{n=0}^3 e_n x_n,
\end{equation}
where $x_0=t,\ (x_1,x_2,x_3) = \bold r$.

A biquaternion whose measure is 1 is called \textit{unitary}.

The operations of conjugation and complex conjugation applied to the product of biquaternions change the order of the factors:
\begin{equation}
\overline{\mathscr{B}_1 \mathscr{B}_2} \ = \overline{\mathscr{B}_2} \ \overline{\mathscr{B}_1}
\end{equation}
\begin{equation}
(\mathscr{B}_1 \mathscr{B}_2)^{\ast} \ = \mathscr{B}_2^{\ast} \ \mathscr{B}_1^{\ast}
\end{equation}

Two biquaternions $\mathscr{B}_1$ and $\mathscr{B}_2$ are \textit{orthogonal} if $Sc(\mathscr{B}_1 \mathscr{B}_2)=0$.

%%%%%%%%%%%% Appendix - OG formula
\subsection{Ostrogradsky-Gauss biquaternionic formulas}\label{Appendix_OG}
For quaternion functions $f(Q)$ of the quaternion argument\footnote{Here we use the quaternion representation, which differs from (\ref{quater_def}) by the absence of the factor $i$ in front of the vector part.}
$Q = (t,\bold r)$, continuously differentiable with respect to each of the coordinates $t,x,y,z$
inside some region $V_4$ of pseudo-Euclidean space and on its boundary $S_3 = \partial V_4$ , formulas of the Ostrogradsky-Gauss type \cite{Deavours},[14] hold:
\begin{equation}
\label{OG_kvat}
  \oint_{S_3} f \ dQ = \int_{V_4} f \hat{D} \ dV_4
\end{equation}
\begin{equation}
\label{OG_kvat2}
  \oint_{S_3} dQ \ f = \int_{V_4} \hat{D} f\ d V_4
\end{equation}
where $\hat{D} = \frac{\partial}{\partial t} + \bold i \frac{\partial}{\partial x} + \bold j \frac{\partial}{\partial y} + \bold k \frac{\partial}{\partial z} $,
$\bold i, \bold j, \bold k$ - basis quaternios, $Q = t + \bold i x + \bold j y + \bold k z $. In (\ref{OG_kvat}) the operator $\hat{D}$ acts to the left of itself, and in (\ref{OG_kvat2}) - to the right.

In \cite{Imaeda} it is proven that similar formulas hold for biquaternion functions $F$ of a real biquaternion argument $Z$:
\begin{equation}
\label{OG_bikvat}
  \oint_{S_3} F \ dZ = \int_{V_4} (F D) \ dV_4
\end{equation}
\begin{equation}
\label{OG_bikvat2}
  \oint_{S_3} dZ \ F = \int_{V_4} (D F) \ dV_4
\end{equation}
Here $D$ denotes the biquaternion gradient\footnote{Our definition is different from the definition used in the \cite{Imaeda} paper, where $D$ denotes $(\partial_t, -\nabla )$. }: $D = (\partial_t, \nabla )$

Let us show that there is an extension of these formulas to the case of two functions $F(Z), G(Z)$ of a real biquaternion argument $Z$.
\begin{equation}
\label{OG}
  \oint_{S_3} F dZ G = \int_{V_4} ( F D G ) d V_4
\end{equation}
Where
\begin{equation}
\label{OG_form}
(F D G) = (F D) G + F (D G)
\end{equation}

The operator $D$ is expanded in a biquaternion basis (14.27) as: $D = \sum \limits_{k=0}^3 e_k \partial_k$, and the coordinate differential as:
$dZ = \sum\limits_{i=0}^3 e_i d x_i$. The left side of the formula (\ref{OG_bikvat2}) will be expressed as:
\begin{equation}
\label{OG_bikvat3}
  \oint_{S_3} dZ \ F = \oint_{S_3} ( \sum\limits_{i=0}^3 e_i d x_i ) \ F = \oint_{S_3} \sum\limits_{i=0}^3 ( e_i F) d x_i = \oint_{S_3} \sum\limits_{i=0}^3 F_i d x_i
\end{equation}
where defined: $F_i=e_i F$.
For the right side of (\ref{OG_bikvat2}) we get:

%%%% insert 1598
\begin{equation}
\label{OG_bikvat4}
 \int_{V_4} (D F)  \  dV_4  = \int_{V_4} ( \sum \limits_{k=0}^3 e_k \partial_k)  F  \  dV_4 = 
  \int_{V_4}  \sum \limits_{k=0}^3  \partial_k( e_k  F)   \  dV_4 =
  \int_{V_4}  \sum \limits_{k=0}^3  \partial_k F_k   \  dV_4
\end{equation}
Thus (\ref{OG_bikvat2}) can be expressed in the form standard for vector analysis:
\begin{equation}
\label{OG2}
  \oint_{S_3} \sum\limits_{i=0}^3 F_i d x_i = \int_{V_4} \sum \limits_{k=0}^3 \partial_k F_k \ dV_4
\end{equation}
It is important to note that the coefficients $F_i = (s_i, \bold F_i)$ are themselves biquaternions. In this case, generally speaking, the existence of the biquaternionic function $F$ is not required: $F_i=e_i F$,
and it is sufficient only for each of the four functions $F_i$ to be continuously differentiable with respect to each of the coordinates. Indeed, let's consider
formula (\ref{OG2}) in particular for scalar components $s_i$:
\begin{equation}
\label{OG22}
  \oint_{S_3} \sum\limits_{i=0}^3 s_i d x_i = \int_{V_4} \sum \limits_{k=0}^3 \partial_k s_k \ dV_4
\end{equation}
This is the Ostrogradsky-Gauss formula in 4-dimensional Euclidean space for a vector function having components $s_i$.
In a similar way, the formula (\ref{OG2}) is satisfied for each of the vector components $F_i$.

Let's use the same basis expansion technique to prove the formula (\ref{OG}). Its left side:
\begin{equation}
  \oint_{S_3} F dZ G = \oint_{S_3} F ( \sum\limits_{i=0}^3 e_i d x_i ) G = \oint_{S_3} \sum\limits_{i=0}^3 ( F e_i G ) d x_i
= \oint_{S_3} \sum\limits_{i=0}^3 A_k d x_i,
\end{equation}
where $A_k= F e_i G$ is denoted. The formula (\ref{OG2}) is applicable to the last integral:
\begin{equation}
  \oint_{S_3}  \sum\limits_{i=0}^3  A_k  d x_i =   \int_{V_4}  \sum \limits_{k=0}^3  \partial_k A_k   \  dV_4 =  \int_{V_4} (  F D  G )  \  dV_4,
\end{equation}
where we took into account that:
\begin{equation}
\label{OG_form2}
\begin{split}
  \sum\limits_{k=0}^3 \partial_k A_k = \sum\limits_{k=0}^3 \frac{\partial}{\partial {x_k}} ( F e_k G) = \sum\limits_{ k=0}^3 ( \frac{\partial F}{\partial {x_k}} e_k G + F e_k \frac{\partial G}{\partial {x_k}} ) = \\
  = ( \sum\limits_{k=0}^3 \frac{\partial F}{\partial {x_k}} e_k) G + F ( \sum\limits_{k=0}^3 e_k \frac{\partial G}{\partial {x_k}}) = ( F D G )
  \end{split}
\end{equation}
Thus, the formula (\ref{OG}) is proven.

As is known from the theory of differential forms, the Ostrogradsky-Gauss and Stokes formulas are a consequence of the general theorem on the external derivative \cite{Arnold} (p. 154).
A remarkable circumstance is that the biquaternion formula (\ref{OG_bikvat}), applied to vector functions in three-dimensional space, gives both of these formulas simultaneously. If we take $F=\bold F(\bold r)$ in (\ref{OG_bikvat}), then integration over the time coordinate is reduced to multiplication by the corresponding time interval $\Delta t$, and we obtain the following formula:
\begin{equation}
\label{OG_tri}
 \Delta t \oint_{S_2} ( \bold F\cdot d \bold r, i \bold F \times d \bold r )   =  \Delta t \int_{V_3} ( \nabla \cdot \bold F, -i \nabla \times \bold F  )  \  dV_3, 
\end{equation} 
which splits into scalar and vector parts:

\begin{equation}
\label{OG_classic}
\oint_{S_2} \bold F\cdot d \bold s   =  \int_{V_3} (\nabla \cdot \bold F ) \  dV_3 
\end{equation}
\begin{equation}
\label{Stoks}
\oint_{S_2} d \bold s \times \bold F = \int_{V_3}  (\nabla \times \bold F  )  \  dV_3, 
\end{equation}
In the formula (\ref{OG_classic}) we replaced $d \bold r$ with $d \bold s $, because here $d \bold r$ has the usual for vector analysis meaning of the oriented area of the two-dimensional surface $S_2$: $d \bold s = \bold n ds$, where $\bold n$ is the normal to this surface. (\ref{OG_classic}) is the classical Ostrogradsky-Gauss formula. (\ref{Stoks}) is a generalized Stokes formula for three-dimensional volume, which can be shown to be equivalent
the usual Stokes formula for circulation along the boundary of a two-dimensional surface.
%%%% end of insert

\end{fulltext}

%%%%%%%%%%%% BIBLIOGRAPHY %%%%%%%%%%%%%

\hfill

\bgroup\obeylines
\textbf{
References
}
\egroup

\hfill

Сергей Яковлевич Котковский

%
%
%%%%%%% ENGLISH %%%%
\clearpage

\end{document}